# Development of Multi-Platform Control and Instrumentation Communications to Increase Operational Reliability – Application to MAST

D. A. Homfray, K. Deakin, S. Khilar, R. King, D. Payne,
M. R. Simmonds, C. Tame and B. Whitehead

*EURATOM/CCFE Fusion Association, Culham Science Centre, Abingdon, UK*

Improving the reliability and reducing the maintenance time to give increased availability is a key feature of developing control & instrumentation (C&I) systems relevant to future fusion devices such as DEMO and to fusion power plants. Standardising access to the multiple platforms comprising C&I systems on working plant including software that analyses data is one aspect of achieving this. This has been realised on the MAST Neutral Beam Injection system (MNBI) following an extensive upgrade to the C&I, to improve the operational reliability of the neutral beam plant.

Keywords: NBI, MAST, Control & Instrumentation, SCADA, Availability, Reliability, Maintenance, Software Methods

## 1. Introduction

On large complex devices such as MAST, JET, DEMO and ITER system reliability and availability are major issues. The importance of maximising physics operations and limited access and time for maintenance increase the requirement for high availability and reliability. It is therefore important that systems are designed to minimise access to restricted areas and that control equipment makes good use of remote debugging and predictive maintenance facilities. With these issues in mind an extensive review and upgrade of the MAST Neutral Beam Injection system (MNBI) Control and Instrumentation (C&I) was undertaken.

The aims of the upgrade were to improve the reliability of C&I in MNBI operations and to prepare the system for the various scenarios of operation for MAST-U [1] (an extensive upgrade to MAST planned to begin in 2013). Better diagnostic and analysis capabilities on the system were identified as vital tools for improving the reliability and availability of MNBI.

From the review, a complete change in methodology and approach to plant control was undertaken. Subsystems were designed to be more modular, allowing the inclusion of further injectors. Subsystems were opened to allow greater access to data and plant signals from intelligent plant and control, maintenance and analysis software.

Over recent years as much as 22% of lost experimental time during MAST campaigns was attributed to MNBI C&I. This was a result of using a C&I system that was initially designed for a different injector type, equipment reaching the end of its operational life, poor diagnostic capabilities and obsolete components (particularly PLCs).

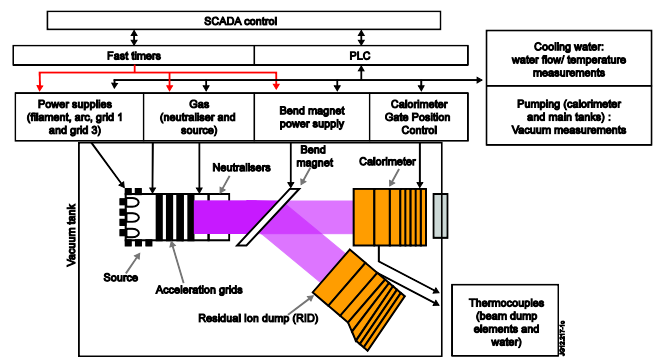

Figure.1 - Schematic of MAST Neutral Beam Injector Control System

Until 2003, MNBI used two duopigatron ion sources on loan from Oak Ridge National Laboratory (ORNL) [2]. These ORNL duopigatron injectors were replaced in 2006 and 2008 as part of the MAST Long Pulse Neutral Beam Upgrade, with JET style PINIs [3]. The upgrade required significant modifications to the MNBI, associated power supplies[4] and interlock systems but only minor improvements were made to the C&I systems.

It became essential that significant improvements were made to increase the C&I system reliability and availability. Implementation began in 2010.

Following the completion of the NB upgrade in March 2011, the MAST M8 Campaign (September 2011 to January 2012) was the most successful campaign to date. Lost experimental time on MAST due to MNBI (including C&I) dropped below 1% (from 22%) and neutral beam had a record breaking campaign providing the highest beam energies recorded.

*author's email: david.homfray@ccfe.ac.uk*

## 2. C&I Upgrade for MAST Neutral Beam

The MNBI plant is controlled and visualised by a Supervisory Control and Data Acquisition system (SCADA), using Programmable Logic Controllers (PLC) and PC based fast timing systems (figure 1). Continuous data is collected showing plant conditions and system health and during an experimental pulse, extra data is collected allowing beam validation.

Following a review of the C&I system a complete change to the methodology in the way MNBI would be operated was adopted. This change defined the way the upgrade was planned, how systems were to be designed and what plant would need to be replaced.

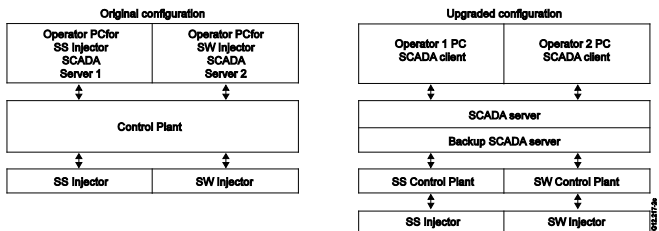

Figure. 2 – Comparison between old and new control methodologies

In the upgrade control and plant subsystems were modularised (figure 2), so each individual injector was standalone, allowing one injector to continue to operate if the other failed during a pulse. This was unlike the previous system of an amalgamated plant structure and opened the possibility of allowing further injectors, designated for MAST-U, to be built from identical subsystems that had already been designed, tested and operated. More importantly with the new methodology, building new infrastructure for MAST-U would leave the current plant untouched, eliminating the potential for the introduction of errors on a working system.

The SCADA system was re-written to allow an operator to control all injectors from one terminal and diagnostic, maintenance and analysis systems were made directly accessible to the operator on the controlling PC. The original suite of software which resided over a number of PC's including several SCADA servers, was restricted to one main SCADA PC facilitating more robust code control. All code was reviewed and modified so that each injector had subsystems with standardized code blocks

The project also involved upgrading a large number of subsystems such as the control of the calorimeter, control of the Titanium Sublimation Pump (TSP), thermocouple instrumentation, network, analysis software and file archiving. The field wiring (wiring from controller to the specific device) between C&I was also replaced and with the introduction of marshalling cubicles (hub for wiring between controller and many devices) it allowed simplified testing and spare unused cables to connect new subsystems to an injector and the ability to isolate signals at a single location.

Parts of the new subsystems were specifically chosen to be easier to configure, diagnose and connect to the new modular system using industrial communication standards such as Ethernet, OPC or Modbus. This allowed many of the systems to be easily connected remotely without the requirement for local access. The NB upgrade allowed many subsystems to be analysed for the first time and with the new ability to communicate, debug, trend and analyse previously unavailable plant signals, several long standing issues were resolved.

A successful example of how opening communications and data transfer improved the reliability and availability of plant was the upgrade to the Titanium Sublimation Pump (TSP) cubicle, which provides the high speed pumping for the vacuum vessels. The cubicle power supply, which operates the titanium filaments, was prone to regular fuse failure. To replace the fuse required stopping operations and making an entry to the MAST Area. As well as fuse failure it was not uncommon for a titanium filament to break during operation and short to an unused filament in the same chamber (figure 3). The frequency of these filament failures was higher than experienced in other laboratories. Eventually there are not enough filaments left to provide the pumping requirement and it is necessary to replace the filaments, which requires a two week shutdown followed by one or two weeks to restore vacuum conditions and recondition the PINI.

During the 2010 upgrade, the PLC controlling the TSP had a communications module added so that access to all signals was possible instead of only to a few hardwired plant signals as was previously the case. The PLC code was re-written to provide more plant status feedback capabilities, made available by the upgraded communications and the addition of diagnostic, de-bugging and storage software allowed improved understanding of the fault scenarios. Combined with appropriate data storage, trends showing filament usage could be obtained. From the new information it was shown that the fuse failure was down to large initial current spikes from the power supply, considerably higher than could be observed previously, which had the potential of damaging the fuses and filaments.

Electronics were designed and built to physically disconnect the power supply control signals in between pulses and a software soft start and stop was added to the PLC, which drove the power supply controller card, attenuating the current spike in the initial phase protecting both fuse and filament. Also the trending capabilities enabled the comparison of filament usage and vacuum tank pressures before and after gettering, allowing filament lifetime predictive capabilities. This provided the opportunity to plan filament changeover before failures occured.

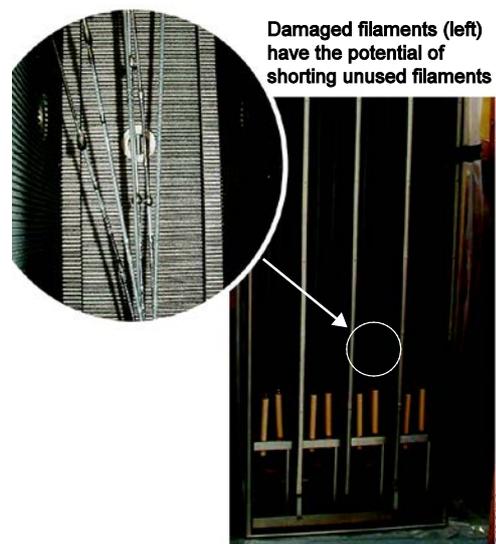

Figure. 3 – Picture of TSP Chamber and melted shorted filaments

In the following MAST M8 Campaign there were no broken filament occurrences instead of the two or three normally seen. The filament lifetime was increased from 12,000s to >35,000s and there was no need for a two week shutdown to replace the filaments, resulting in savings of around £30,000 and several weeks of downtime.

## 3. Multi-Platform Control and Instrumentation Communication Deployment

Reliability and availability on a complex system such as MNBI is improved immeasurably when remote access is available to the plant. This allows monitoring, debugging and fault finding without the need for area entry. The ability to quickly write code which enables recording, analysing and trending, with standard functions calling plant signals directly, that may not be normally of interest can quickly highlight and diagnose system faults.

Multi-platform standardized communications protocols are widely used in industry to allow communication between multiple plant devices from different vendors. This is not normally extended to the use of analysis, and maintenance codes (such as trending of vacuum conditions) running on PCs as it can be difficult to implement and quite specialist. Software used to visualize and clarify plant data is normally a passive, one-way process, where specified data is recorded, a file created and analysis is performed. The results, showing a specific device may need either fine tuning or attention, would normally require human intervention, providing routes for error.

A method allowing all available plant signals (not just the routinely collected signals) to be monitored and acted upon by non real-time control codes (such as analysis, maintenance and/or de-bugging), using standardized code without the need to understand either the device or communication protocols, opens opportunities for a more reliable and available system.

An optimal solution would be to make the SCADA tags (plant signals registered on the SCADA main variable or tag database) directly available to codes for analysis, maintenance and debugging PC's. Unfortunately the main method available for direct access to the SCADA system is through OPC DA (OLE for Process Control - Data Access) which was originally based on Microsoft's OLE COM (Component Object Model) and DCOM (Distributed Component Object Model) technologies and requires each PC to be configured to allow signal access across the network, which is difficult and time consuming. Third party DCOM configuration packages are available but are expensive, difficult to use and require specialist knowledge.

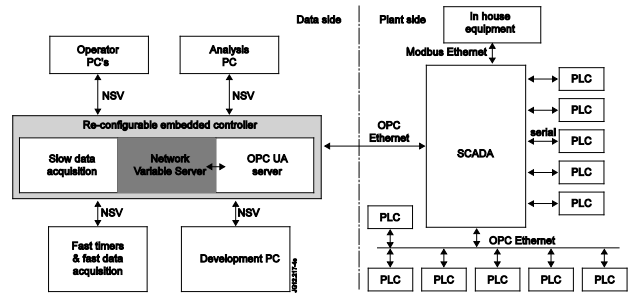

Figure 4 – Overview of MAST NBI Multi-Platform Communication Architecture

A method which allowed access to these SCADA tags without the requirement for configuration of the host PC and without requiring specific knowledge of the plant device or its communication protocol was devised. Figure. 4 shows the method used, where a re-configurable embedded controller already used for collecting slow analogue data (thermocouple data at ~100Hz) was re-programmed (providing both data recording and server parallel process loops) to connect the SCADA tags to a vendor proprietary network variable server (National Instruments network variables) which is easily accessed by simple function calls in VB, C and LabVIEW. The software interface between the SCADA System and the network variable server uses the modern OPC UA (Unified Architecture) server architecture which allows a secure and reliable cross platform framework for non-Microsoft systems. The OPC UA framework allows the controller to be added to the SCADA system driver database like a PLC and the network variables to look like signals from that PLC.

The software on start-up automatically recognises all Network Variables (NV) added to the embedded controller, separates data type and links them to standard OPC format variable name for each variable. It creates an OPC UA Server and then using two separate processes, polls for changes on the NV side or for an event change on the SCADA tag side and updates the other side of the link. The polling time on the NV side is currently around 200ms and therefore this centralised platform is not intended for mission critical signals but for general data movement between pulses and slow data collection during operations.

Writing software to read or write plant signals, not normally recorded, but available to SCADA either through PLC or other intelligent controllers becomes a formality. To register a plant signal, a NV name is added to the embedded controller through a standard interface, the variable is automatically registered by the SCADA driver database (database defining how the SCADA communicates with the specific plant signal). A tag is then written on the SCADA main tag database where the desired plant signal is linked to the registered signal on the SCADA driver database.

This method not only allows for multiple intelligent controllers and software suites to access any of the data available to the embedded controller or SCADA system but it also enables multiple systems to monitor a single variable eliminating the possibility of time propagation errors from multiple cloned variables which can lead to unexpected errors and prove difficult to fault find.

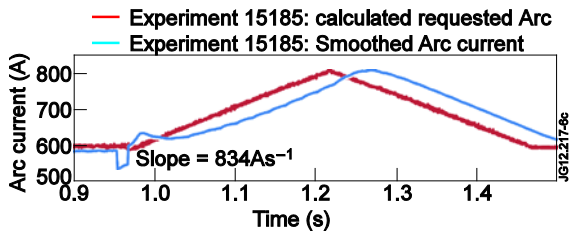

Figure 5 – Graph of Calculated Arc Current Request and Measured

Some examples of how this cross platform methodology has improved reliability and availability:

1. The MNBI Operator cannot setup plant for the next experiment until all analysis is performed (SCADA independent analysis software). The operator's user interface is told if analysis is successful and if it is safe to proceed.
2. After the analysis has been completed, pulse information and alarms are available to the operator directly from the analysis PC, without the need for separate data visualization code.
3. A single variable for pulse number and pulse type allows the analysis software to omit certain functions if certain data sets are inappropriate.
4. Remote resets and on/off capabilities from the operator user interface, for subsystems on the data or plant side without the requirement of area access.

## 4. Future Implementation

An investigation to see if the scan rate of the OPC UA Server can be increased is underway. Currently it is not possible to use event handlers on the NV side of the embedded controller and therefore improvement in efficiencies in polling are required. One of the severe limitations is that a connection needs to be made to the NV side for each data type and then closed before the next data type is checked. This could be made more efficient if multiple connections to each variable type could be kept open, all collated and converted to a variant type variable (data type that can be used to represent any other data type) and then sent to update the OPC UA Server.

Failure of small power supplies used in euroracks to power small control electronics happen frequently and are time consuming to find. Design work is currently underway for an electronics card to be installed in each cubicle, to monitor power supply status, using Modbus Ethernet. Separate diagnostic codes (reducing resource load on the SCADA system) could then be utilised to check for a failing or failed plant providing the operator with the location allowing immediate replacement.

For MAST-U, a real time NB power control system to actively control beam timing and power is currently being developed. This will involve an in-house built FPGA-based master [5] (connected to the SCADA System using Modbus) controlling individual injector based FPGA controllers, which will include controlling beam power by actively controlling arc voltage and current [6]. From initial measurements it can be seen there are associated time lags due to filament thermal effects, the arc notch (causing a small overshoot at beginning of the beam) and the arc controller feedback circuits (figure 5). The addition of a 'pre-emphasis' signal (additional component to the main signal) to the requested arc waveform to achieve the desired actual arc current is being considered. It is hoped that a separate analysis program can be used to automatically fine tune the arc request for different beam voltage and power.

## 5. Conclusions

An extensive upgrade to the MNBI C&I system was implemented to improve reliability, availability and prepare for MAST-U. Lost experimental time on MAST due to MNBI dropped from 22% to below 1% as a result, allowing for one of the most successful MAST campaigns with the highest neutral beam energies achieved to date.

A new network variable server methodology was implemented, enabling better access to plant from software using standardised functions allowing quicker and easier code to be produced without specialised knowledge about the plant or the communication protocols. The increased analysis, debug and maintenance capability enables better performance monitoring and trending and therefore increases reliability and availability.

## Acknowledgments

This work was funded by the RCUK Energy Programme under grant EP/I501045 and the European Communities under the contract of Association between EURATOM and CCFE. The views and opinions expressed herein do not necessarily reflect those of the European Commission